\documentclass[aps,floats]{revtex4}
\usepackage{amsmath,amssymb}
\usepackage{graphicx,epsfig}
\usepackage[greek,english]{babel}

\begin{document}
\bibliographystyle {plain}

\def\oppropto{\mathop{\propto}} 
\def\opsimeq{\mathop{\simeq}}
\def\opoverderline{\mathop{\overline}}
\def\operarrow{\mathop{\longrightarrow}}
\def\opsim{\mathop{\sim}}

\def\fig#1#2{\includegraphics[height=#1]{#2}}
\def\figx#1#2{\includegraphics[width=#1]{#2}}


\title{ Statistical properties of the Green function in finite size \\
for Anderson Localization models with multifractal eigenvectors  } 


\author{ C\'ecile Monthus }
 \affiliation{Institut de Physique Th\'{e}orique, 
Universit\'e Paris Saclay, CNRS, CEA,
91191 Gif-sur-Yvette, France}

\begin{abstract}
For Anderson Localization models with multifractal eigenvectors on disordered samples containing $N$ sites, we analyze in a unified framework the consequences for the statistical properties of the Green function. We focus in particular on the imaginary part of the Green function at coinciding points $G^I_{xx}(E-i \eta)$ and study the scaling with the size $N$ of the moments of arbitrary indices $q$ when the broadening follows the scaling $\eta=\frac{c}{N^{\delta}}$. For the standard scaling regime $\delta=1$, we find in the two limits $c \ll 1$ and $c \gg 1$ that the moments are governed by the anomalous exponents $\Delta(q)$ of individual eigenfunctions, without the assumption of strong correlations between the weights of consecutive eigenstates at the same point. For the non-standard scaling regimes $0<\delta<1$, we obtain that the imaginary Green function follows some Fr\'echet distribution in the typical region, while rare events are important to obtain the scaling of the moments. We describe the application to the case of Gaussian multifractality and to the case of linear multifractality.

\end{abstract}

\maketitle

\section{ Introduction }

For Anderson Localization models \cite{50years} defined on $N$ sites,
or for $N \times N$ random matrices, 
the statistics of the $N$ eigenvalues $E_n$ and 
of the corresponding eigenvectors $\vert \phi_n>$
\begin{eqnarray}
H = \sum_{n=1}^N E_n \vert \phi_n><\phi_n \vert
\label{hspectral}
\end{eqnarray}
have been much studied,
 in particular at Anderson transitions where eigenvectors display multifractality
(see the reviews \cite{janssenrevue,mirlin_revue2000,mirlin_revue}
and references therein). Besides short-ranged Anderson models in finite dimension $d>2$, where multifractality occurs only at the critical point between the exponentially localized phase and the delocalized ergodic phase, there exists other 
Anderson models where multifractality appears also outside the critical point.
In the localized phase, multifractality occurs whenever localized eigenfunctions 
are only power-law localized with respect to the size $N$ of the Hilbert space 
(see more details in the introduction of \cite{c_levyloc}) :
examples are (i)  Anderson models in finite dimension $d$ with power-law hoppings 
(ii)  nearest-neighbor Anderson models on trees or on other spaces of effective infinite dimensionality where the Hilbert space grows exponentially with the linear length
(iii) some matrix models where the matrix elements are rescaled with some power of the size $N$ of the matrix.
In the delocalized phase, the existence
 of a non-ergodic phase displaying multifractality has been
 much debated recently,
in particular for the short-ranged Anderson model
 on the Bethe lattice, either with boundaries \cite{us_cayley,mirlin_tik}
or without boundaries \cite{biroli_nonergo,luca,mirlin_ergo,altshuler_nonergo,lemarie,ioffe}, and for random matrix models, like the Generalized-Rosenzweig-Potter (GRP) Matrix model \cite{kravtsov_rosen,biroli_rosen}, 
or the L\'evy Matrix model \cite{cizeau,biroli_levy,c_levygolden}.

Whenever eigenfunctions display multifractality, it is important to understand
 the consequences for the statistics of the 
 Green function defined for the complex variable $z=E-i \eta$ 
\begin{eqnarray}
G_{xy}(z=E-i \eta) \equiv <x \vert \frac{1}{z-H}  \vert y >
 =  \sum_{n=1}^N \frac{ <x \vert \phi_n> < \phi_n \vert y > } { (E-E_n) -i \eta}
\label{gspectral}
\end{eqnarray}
The Green function is the basic object from which the physical observables can be obtained
(see the reviews \cite{janssenrevue,mirlin_revue2000,mirlin_revue}).
The observable that has been the most discussed is the imaginary part
at coinciding points $x=y$, because it leads to the local density of states $\rho_x(E)$
which can be considered as an order parameter for the Anderson transition (see the reviews \cite{janssenrevue,mirlin_revue2000,mirlin_revue})
\begin{eqnarray}
\rho_x(E) \equiv   \sum_{n=1}^N \phi^2_n(x)  \delta(E-E_n) 
\label{rhoxe}
\end{eqnarray}
in the limit $\eta \to 0$
\begin{eqnarray}
G^I_{xx}(z=E-i \eta) =   \sum_{n=1}^N \phi^2_n(x)  
\frac{  \eta} { (E-E_n)^2 + \eta^2 }  \opsimeq_{\eta \to 0} 
\pi  \rho_x(E)
\label{gixx}
\end{eqnarray}
The Green function for two different points $x \ne y$ contains the information on the Landauer two-point transmission
where an incoming and an outgoing wires are attached to the two points $x$ and $y$
in order to probe the conductance of the sample \cite{janssen_transmission,us_twopoint,us_aoki}. 
The direct relation that has been found between this Landauer two-point transmission and the product of two local densities of states \cite{janssen_transmission} shows the central role played by
the local density of states, or equivalently by the imaginary part of the Green function at coinciding points of Eq \ref{gixx}
on which we will mostly focus in the present paper.

However since one is interested
 in the double limit $N \to +\infty$ and $\eta \to 0$,
it is essential to state what is the scaling relation between the two variables
in terms of some exponent $\delta>0$ \cite{mirlin_tik,biroli_rosen,ioffe}
\begin{eqnarray}
\eta = \frac{c}{N^{\delta}}
\label{etadelta}
\end{eqnarray}
In the standard scaling regime $\delta=1$ \cite{mirlin_tik,biroli_rosen,ioffe},
 where the broadening $\eta= \frac{c}{N}$ scales as
the level spacing
\begin{eqnarray}
\Delta = \frac{1}{N \rho(E)}
\label{levelspacing}
\end{eqnarray}
(where $\rho(E)= \frac{1}{N} \sum_{n=1}^N  \delta(E-E_n) $
 is the global density of states),
the local density of states 
is expected to inherit the statistical properties of
a single eigenfunction $\phi^2_E(x) $ of energy $E$
 \cite{janssenrevue,mirlin_revue2000,mirlin_revue}
\begin{eqnarray}
G^I_{xx}(z=E-i \frac{c}{N} ) \opsimeq_{law} \pi N \phi^2_E(x)  
\label{ldos1state}
\end{eqnarray}
although this relation has been recently found to be violated for non-integer moments of indices $0<q<\frac{1}{2}$ for the Anderson Localization model on the Cayley tree \cite{mirlin_tik}. A natural question is whether this violation is very specific
to the tree geometry with boundaries or whether it could also happen in other cases. 
In addition, three recent studies \cite{mirlin_tik,biroli_rosen,ioffe} have found that 
the characterization of non-ergodic delocalized phases requires to analyze 
the non-standard scaling of Eq. \ref{etadelta} with an exponent $0<\delta<1$,
so that the broadening $\eta$ is much bigger than the level spacing $\Delta$
and contains a sub-extensive number of levels 
\begin{eqnarray}
M_{av} = \frac{\eta}{\Delta} =c \rho(E) N^{1-\delta}
\label{mavdelta}
\end{eqnarray}
The main idea is that the standard scaling $\delta=1 $ is enough to characterize the delocalized ergodic phase,
but is not enough to discriminate between the localized phase and some delocalized non-ergodic phase \cite{mirlin_tik,biroli_rosen,ioffe}.
We refer the reader to the three works \cite{mirlin_tik,biroli_rosen,ioffe} for the various points of view
that lead to the introduction of an anomalous scaling exponent $0<\delta<1$, and for the concrete applications
to the Anderson model on the Cayley tree or the Bethe lattice, and the Generalized Rosenzweig-Porter random matrix model.

The aim of the present work is to re-analyze in a unified framework
the general statistical properties of the Green function
in finite size $N$
as a function of the broadening scaling $\eta=c/N^{\delta}$
whenever the eigenfunctions display multifractality.
 To simplify the discussion, the amplitudes are assumed to be real variables
$\phi_n(x)=<x \vert \phi_n>=<\phi_n \vert x> $,
so that the real part and the imaginary part of the Green function
 of Eq. \ref{gspectral} read respectively
\begin{eqnarray}
G^R_{xy}(z=E-i \eta) && =\sum_{n=1}^N \phi_n(x) \phi_n(y)
\frac{ (E-E_n) } { (E-E_n)^2 + \eta^2 }  
\nonumber \\
 G^I_{xy}(z=E-i \eta) &&\equiv   \sum_{n=1}^N \phi_n(x) \phi_n(y) 
\frac{  \eta} { (E-E_n)^2 + \eta^2 }  
\label{grgi}
\end{eqnarray}
The analysis of the statistics of the Green function will be based on the structure of this spectral decomposition involving a sum over $N$ terms,
where the eigenvalues $E_n$ and the multifractal eigenstates $\phi_n$ appear.

The paper is organized as follows.
Section \ref{sec_multif} contains a reminder on the multifractal formalism for eigenfunctions.
Section \ref{sec_real} is devoted to the statistics of the real part $G^R_{xy}(E)$
for $\eta=0$, both at coinciding points $x=y$ and at different points $x \ne y$ :
the emphasis is on the heavy-tails that appear and on the system-size dependence
of their amplitudes. 
We then turn to the statistics of the imaginary part $G^I_{xx}(E-i \eta)$
in section \ref{sec_imaginary} and explain the role of the scaling
of the broadening $\eta=\frac{c}{N^{\delta}}$ :
 the standard scaling $\delta=1$ is then studied in the limit $c \ll 1$
in section \ref{sec_onelevel}
 and in the limit $c \gg 1$ in section \ref{sec_many},
while the non-standard scaling $0<\delta<1$
 is analyzed in section \ref{sec_delta}.
Our conclusions are summarized in section \ref{sec_conclusion}.

\section { Reminder on the multifractal statistics of eigenfunctions }

\label{sec_multif}

After its introduction in the field of turbulence 
(see the book \cite{frisch} and references therein),
the notion of multifractality
 has turned out to be relevant in many areas of physics
(see for instance \cite{halsey,Pal_Vul,Stan_Mea,Aha,Meakin,harte,duplantier_houches}),
in particular at critical points of random classical spin models
\cite{Ludwig,Jac_Car,Ols_You,Cha_Ber,Cha_Ber_Sh,PCBI,BCrevue,Thi_Hil,cnancy}.
More recently, multifractality has been used to analyze the
ground state wavefunction in pure quantum spin models
\cite{jms2009,jms2010,jms2011,moore,grassberger,atas_short,atas_long,luitz_short,luitz_o3,luitz_spectro,luitz_qmc,jms2014,alcaraz,c_gsmultif}, 
and in the field of Many-Body-Localization models
 \cite{luca_mbl,alet,fradkin,santos,c_mblentropy,c_mblrgeigen}.
In this section, we recall the usual multifractal properties for Anderson Localization models \cite{janssenrevue,mirlin_revue2000,mirlin_revue}
that will be useful in the present paper : 
we stress that we consider only the usual framework where all sites have the same statistical properties.
In particular, this excludes the case of the Anderson Localization model on
the tree geometry with boundaries where the sites have different properties as a function
of their positions between the root and the boundaries.

\subsection{ Multifractal exponents $\tau(q)= (q-1)D(q)=(q-1)+\Delta(q)$ }

The weight
\begin{eqnarray}
w_n \equiv  \vert < x \vert \phi_n> \vert^2
\label{wn}
\end{eqnarray}
of the eigenvector $\vert \phi_n> $ on the site $x$
has to satisfy the normalization of the eigenfunction 
\begin{eqnarray}
\sum_{x=1}^N \vert < x \vert \phi_n> \vert^2 =1 
\label{normeigen}
\end{eqnarray}
As a consequence, the weight $w_n$ 
is a random variable distributed with some 
normalized probability distribution $\Pi_N(w)$ defined on $0 \leq w \leq 1$
\begin{eqnarray}
\int_0^1 dw \Pi_N(w)  = 1
\label{wnnorma}
\end{eqnarray}
and the first moment is fixed to be
\begin{eqnarray}
\int_0^1 dw \Pi_N(w) w = \frac{1}{N}
\label{wnav}
\end{eqnarray}

A convenient way to characterize the statistics of $w$ is 
via the scaling with the size $N$ of the moments of arbitrary index $q>0$
\begin{eqnarray}
Y_q(N) \equiv N \int_0^1 dw \Pi_N(w) w^q \propto N^{-\tau(q)}  
\label{yqtauq}
\end{eqnarray}
Since delocalized eigenfunctions are characterized by the single scaling $w \propto \frac{1}{N}$ for all weights and correspond to the exponents
\begin{eqnarray}
Y^{deloc}_q(N)  \propto N^{1-q}  
\label{yqdeloc}
\end{eqnarray}
it is useful to introduce the generalized fractal dimensions
 \begin{eqnarray}
D(q) = \frac{ \tau(q)}{q-1 }
\label{Dqtauq}
\end{eqnarray}
It is also convenient to introduce the anomalous exponents
 \begin{eqnarray}
\Delta(q) =  \tau(q)-(q-1)
\label{Deltaqtauq}
\end{eqnarray}
that are expected to govern the scaling of the moments of the local density of states,
if one assumes that Eq. \ref{ldos1state} is valid,
as usually done \cite{janssenrevue,mirlin_revue2000,mirlin_revue}
\begin{eqnarray}
\overline{ [\rho_x(E)]^q } \propto \overline{ [ N \phi^2_E(x)  ]^q } \propto N^{q-1-\tau(q) } = N^{-\Delta(q)}
\label{ldos1statedeltaq}
\end{eqnarray}

\subsection{ Multifractal spectrum $f(\alpha)$ }

The probability distribution $\pi_N(\alpha)$ of the exponent 
\begin{eqnarray}
\alpha \equiv - \frac{\ln w}{\ln N}
\label{alphadef}
\end{eqnarray}
is obtained from $\Pi_N(w)$  by the change of variable $w=N^{-\alpha}$
\begin{eqnarray}
\pi_N(\alpha) = (\ln N) N^{-\alpha} \Pi_N(N^{-\alpha} )
\label{pialpha}
\end{eqnarray}
The multifractal spectrum $f(\alpha)$ is defined by the scaling of the leading power-law behavior in $N$
\begin{eqnarray}
\pi_N(\alpha) \propto N^{f(\alpha)-1}
\label{falpha}
\end{eqnarray}
Its maximum unity corresponds to the typical exponent $\alpha_0$
\begin{eqnarray}
f(\alpha_0) =1
\label{falpha0unity}
\end{eqnarray}
All others values $\alpha \ne \alpha_0$ satisfying $f(\alpha)<1$
occur with probabilities that decay with $N$ as the power-laws $N^{-(1-f(\alpha))}$.
When one consider the $N$ weights of a given sample,
the smallest exponent $\alpha_{smallest}$ and the biggest exponent $\alpha_{biggest}$
 that can be seen are the values occuring with a probability of order $\frac{1}{N}$ corresponding to the values where the multifractal spectrum vanishes
\begin{eqnarray}
f(\alpha_{smallest}) = 0 =f(\alpha_{biggest}) 
\label{falphaminmax}
\end{eqnarray}

The saddle-point calculus in $\alpha$ of the moments of Eq. \ref{yqtauq}
\begin{eqnarray}
Y_q(N) \equiv N \int_0^1 dw \Pi_N(w) w^q
 = N \int d\alpha \pi_N(\alpha) N^{-\alpha q} \propto \int d\alpha N^{f(\alpha)-\alpha q} 
\label{yqsaddle}
\end{eqnarray}
yields that it is dominated by some saddle-point value $\alpha_q$
\begin{eqnarray}
 - \tau(q) = {\rm max}_{\alpha} \left( f(\alpha) - q \alpha \right)= f(\alpha_q)-q \alpha_q
\label{legendrec}
\end{eqnarray}
so that $\tau(q)$ and $f(\alpha)$ are related via the Legendre transform
\begin{eqnarray}
   \tau(q)+f(\alpha) && = q \alpha 
\nonumber \\
   \tau'(q) && =  \alpha 
\nonumber \\
f'(\alpha) && = q 
\label{legendre}
\end{eqnarray}

This means that each moment of index $q>0$
 is a priori dominated by a different value $\alpha_q $,
that optimizes the rare probability $N^{f(\alpha_q)-1}$ that decay with $N$ 
for any $\alpha_q \ne \alpha_0$, with the contribution $w^{-\alpha q}$
to the moment of order $q$. 
It is useful to keep in mind the following special values :

(i) for $q=0$, the normalization of Eq. \ref{wnnorma} 
means that $Y_0(N)=N$ 
corresponding to $\tau(0)=-1$ and to $f(\alpha_0)=1$,
where 
\begin{eqnarray}
   \alpha_0 = -  \int_0^1 dw \Pi_N(w) \frac{ \ln w }{\ln N}
\label{alphazero}
\end{eqnarray}
 represents the typical exponent (Eq. \ref{falpha0unity}).

(ii) for $q=1$, the scaling of the first moment (Eq \ref{wnav})
means that $Y_1(N)=N$ corresponding to $\tau(q=1)=0$ and
\begin{eqnarray}
   \alpha_1   = f(\alpha_1) = D(1)
\label{alphaun}
\end{eqnarray}
is called the 'information dimension'.

(iii) for $q \to +\infty$, $Y_q(N)$ is dominated by the minimal exponent
introduced in Eq. \ref{falpha0unity}
\begin{eqnarray}
   \alpha_{+\infty} = \alpha_{smallest}  = D(+\infty)
\label{alphaqinfty}
\end{eqnarray}

To be more concrete, it is now useful to mention two simples examples.

\subsection { Example of the log-normal distribution for the weights }

\label{sec_lognormal}

The log-normal distribution of the weight $w$
\begin{eqnarray}
\Pi_N(w) \simeq \frac{1}{ w \sqrt{2 \pi \sigma^2 \ln N } }  e^{- \frac{ (\ln w +\alpha_0 \ln N)^2 }{2  \sigma^2 \ln N }}
\label{lognormal}
\end{eqnarray}
corresponds to the Gaussian distribution of the exponent $\alpha$ (Eq \ref{pialpha})
\begin{eqnarray}
\pi(\alpha) = (\ln N) N^{-\alpha} \Pi_N(N^{-\alpha} ) = 
\sqrt{ \frac{  \ln N }{ 2 \pi \sigma^2   } } e^{- \frac{ (\alpha -\alpha_0 )^2 }{2  \sigma^2  } \ln N}
= \sqrt{ \frac{  \ln N }{ 2 \pi \sigma^2   } } N^{- \frac{ (\alpha -\alpha_0 )^2 }{2  \sigma^2  } }
\label{pialphalogn}
\end{eqnarray}
so that the multifractal spectrum of Eq. \ref{falpha} is quadratic
\begin{eqnarray}
f(\alpha) = 1 - \frac{ (\alpha -\alpha_0 )^2 }{2  \sigma^2  } 
\label{falphalogn}
\end{eqnarray}
The minimal exponent $\alpha_{min}$ and the maximal exponent $\alpha_{max} $
of Eq. \ref{falphaminmax}
\begin{eqnarray}
\alpha_{smallest} && = \alpha_0 -\sqrt{ 2  \sigma^2  } 
\nonumber \\
\alpha_{biggest} && = \alpha_0 +\sqrt{ 2  \sigma^2  } 
\label{alphaminmaxlogn}
\end{eqnarray}

The moments of arbitrary index $q>0$ reads
\begin{eqnarray}
Y_q(N) = N \int_0^1 dw  \Pi_N(w)  w^q  && = 
N \int_{0}^{+\infty } d\alpha  \sqrt{ \frac{  \ln N }{ 2 \pi \sigma^2   } } 
 e^{- \frac{ (\alpha -\alpha_0 )^2 }{2  \sigma^2  } \ln N} e^{-\alpha q \ln N}
 = N^{1+\frac{q^2 \sigma^2}{2} -\alpha_0 q}
\label{wqmbl}
\end{eqnarray}
The condition on the first moment (Eq \ref{wnav}) fixes the parameter
\begin{eqnarray}
\sigma^2 = 2( \alpha_0 -1) 
\label{sigmaln}
\end{eqnarray}
so that Eq. \ref{wqmbl} becomes
\begin{eqnarray}
Y_q(N) 
 = N^{(1-q) [1-(\alpha_0 -1) q]}
\label{yqgauss}
\end{eqnarray}
corresponding to the generalized dimensions (Eq. \ref{Dqtauq})
\begin{eqnarray}
D(q) =1-(\alpha_0 -1) q
\label{dqgauss}
\end{eqnarray}

\subsection { Example of the truncated power-law distribution for the weights}

\label{sec_levy}

The case of a truncated power-law of L\'evy index $0<\nu<1$
\begin{eqnarray}
 \Pi_N(w)  = \frac{(1-\nu) }
{N w^{1+\nu} } \theta ( a N^{-\frac{1}{\nu}}  \leq  w \leq 1 )
\label{wlevynu}
\end{eqnarray}
where
\begin{eqnarray}
a^{\nu} =  \frac{ 1-\nu  }{\nu}
\label{alevynu}
\end{eqnarray}
corresponds for the exponent $\alpha$ to the distribution (Eq \ref{pialpha})
\begin{eqnarray}
\pi_{N}(\alpha) = (\ln N) N^{-\alpha} \Pi_N(N^{-\alpha} ) = c  (\ln N) N^{\alpha \nu -1} 
 \theta \left(  0 \leq \alpha \leq \frac{1}{\nu} - \frac{\ln a}{ \ln N}  \right)
\label{alphalev}
\end{eqnarray}
so that the multifractal spectrum of Eq. \ref{falpha} is linear
\begin{eqnarray}
f(\alpha) =  \alpha \nu  
\  \theta \left(  0 \leq \alpha \leq \frac{1}{\nu}  \right)
\label{falphalev}
\end{eqnarray}
with the typical exponent $\alpha_0=\frac{1}{\nu} $
and the smallest exponent $\alpha_{smallest}=0$.

The behavior of the moments
\begin{eqnarray}
Y_q(N) =N \int_0^1 dw \Pi_N(w) w^q = (1-\nu) \int_{ a N^{-\frac{1}{\nu}}}^1 dw  w^{q-1-\nu}
=  (1-\nu)  \frac{ 1 -a^{q-\nu} N^{\frac{\nu-q}{\nu}} }{q-\nu}
\label{wnqlev}
\end{eqnarray}
depend on the sign of $(q-\nu)$ :

(i) for $q> \nu$, they remain finite
\begin{eqnarray}
Y_{q>\nu}(N) \opsimeq_{N \to +\infty}  \frac{ 1-\nu }{q-\nu}
\label{wnqlevbig}
\end{eqnarray}

(ii) for $ q<\nu$, they diverge as
\begin{eqnarray}
Y_{ q<\nu}(N) \simeq  (1-\nu) \frac{ a^{q-\nu} N^{1-\frac{q}{\nu}} }{\nu-q}
= \frac{\nu}{\nu-q} a^q N^{1-\frac{q}{\nu}}
\label{wnqlevsmall}
\end{eqnarray}

So the generalized fractal dimensions $D(q)$ de Eq. \ref{Dqtauq} read
\begin{eqnarray}
D(q) && = \frac{1-\frac{q}{\nu}} {1-q} \ \ {\rm for }  \ \ q \leq \nu
\nonumber \\
 D(q) && = 0 \ \ \ \ \ \ \ \  {\rm for } \ \  q \geq \nu
\label{dqlevy}
\end{eqnarray}

For $\nu=\frac{1}{2}$, this spectrum is well-known as the 'Strong Multifractality' critical spectrum \cite{mirlin_fyodorov,mirlin_four} :
it appears in particular at Anderson Localization Transition
in the limit of infinite dimension $d \to +\infty$ \cite{mirlin_revue}
or in long-ranged power-law hoppings in one-dimension 
 \cite{levitov1,levitov2,levitov3,levitov4,mirlin_evers,fyodorov,
fyodorovrigorous,oleg1,oleg2,oleg3,oleg4,oleg5,oleg6,olivier_per,olivier_strong,olivier_conjecture,us_strongmultif},
 in the generalized Rosenzweig-Potter matrix model of \cite{kravtsov_rosen}, and in the L\'evy Matrix Model \cite{c_levyloc}.
Recently, it has been also found for the Many-Body-Localization transition \cite{c_mblrgeigen,c_mblentropy}.

For $\nu <\frac{1}{2}$, Eq \ref{dqlevy} describes the multifractal properties of the localized phase of some of these models,
as found in \cite{kravtsov_rosen,c_levyloc,c_mblrgeigen,c_mblentropy}.

\subsection{Multifractality and correlations  } 

Whenever the weight $w_n(x)=\phi^2_n(x)$ of the eigenfunction $\phi_n$ at sites $x$
follows some multifractal statistics as described above, one expects that there exists
 some correlations between the weights involving different eigenfunctions and/or different sites \cite{janssenrevue,mirlin_revue2000,mirlin_revue}.
However for each given observable involving different weights, the important question in practice is
 whether these correlations need to be taken into account to evaluate the dominant scaling or not.
Let us consider two explicit examples of observables.

A first example concerns the Inverse Participation Ratios (I.P.R.)  
\begin{eqnarray}
I_q (N) \equiv \sum_{x=1}^N [\phi^2_n(x) ]^q
\label{Iq}
\end{eqnarray}
that involve the weights of a given eigenstate $\phi_n$ on all the sites $x=1,..,N$ of the disordered sample. 
In the multifractal literature \cite{janssenrevue,mirlin_revue2000,mirlin_revue},
these I.P.R. are governed by the same multifractal exponents $\tau(q)$ introduced above for the statistics of a single weight 
\begin{eqnarray}
I_q(N) \propto N^{-\tau(q)}  \propto Y_q(N) \equiv N \int_0^1 dw \Pi_N(w) w^q 
\label{wnqdqipr}
\end{eqnarray}
This means that in the I.P.R. of Eq. \ref{Iq}, the spatial correlations existing between the weights $\phi^2_n(x) $ on different sites $x$ 
actually do not change the scaling of Eq. \ref{wnqdqipr} that corresponds to a computation neglecting these correlations.

Another example is the local density of states of Eq. \ref{rhoxe} or the imaginary part of the Green function at coinciding points (Eq \ref{gixx})
that involve the weights $ \phi^2_n(x)$ of the $N$ eigenstates $\phi_n$ with $n=1,..,N$ on the same site $x$.
Since multifractality means 'sparsity', two independent multifractal 
eigenfunctions are not expected to 'see' each other.
 However at standard Anderson transitions where there is level repulsion,
the multifractal consecutive eigenstates have to be strongly correlated in order to 'see' each other and to produce level repulsion. 
On the contrary, when multifractality occurs within the localized phase (see the introduction), there is no level repulsion (Poisson statistics)
and the multifractal consecutive eigenstates do not 'see' each other
 and are not expected to be strongly correlated.
This raises the question whether the standard assumption of Eq. \ref{ldos1state}
is valid only in the presence of very strong correlations between the weights of consecutive eigenstates when there is level repulsion,
but is violated otherwise : see the recent discussion for the Anderson Localization model on the Cayley tree \cite{mirlin_tik}.
This question on the validity of Eq. \ref{ldos1state} in general Anderson Localization models with multifractal eigenfunctions
has been one of the motivation of the present work.

\section{ Probability distribution of the real Green function $G^R(E)$   }

\label{sec_real}

In this section, we consider the case $\eta=0$ and discuss the statistical properties of the real Green function (Eq. \ref{grgi})
\begin{eqnarray}
G^R_{xy}(E) && =\sum_{n=1}^N 
\frac{ \phi_n(x) \phi_n(y)  } { E-E_n }  
\label{grxyeta0}
\end{eqnarray}
first for coinciding points $x=y$ and then for different points $x \ne y$.

\subsection { Heavy tail analysis for the probability distribution of $G^R_{xx}$  }

The real Green function of Eq. \ref{grxyeta0} for $x=y$
\begin{eqnarray}
G^R_{xx}(E) && =\sum_{n=1}^N 
\frac{ \phi^2_n(x)   } { E-E_n }  
\label{grxxeta0}
\end{eqnarray}
may become arbitrarily large $\vert G^R \vert  \to +\infty$
only if the external energy $E$ becomes very close $s \to 0$ to the 
nearest energy-level $E_{n_E}$ 
\begin{eqnarray}
\vert E- E_{n_E} \vert = s \Delta
\label{diffe}
\end{eqnarray}
where $\Delta=\frac{1}{N \rho(E)}$ is the level spacing of Eq. \ref{levelspacing} and 
where $s$ is an $O(1)$ random variable distributed exponentially
\begin{eqnarray}
R(s) = e^{-s}
\label{rsexp}
\end{eqnarray}

Then the corresponding term dominates over the others in Eq. \ref{grxxeta0},
and the Green function can be approximated by this biggest term
\begin{eqnarray}
\vert G^R_{xx}(E)  \vert &&   \opsimeq_{ s \to 0}  \frac{ w  }{ \Delta s  } = \frac{ N \rho(E) w  }{  s  } 
\label{grxxssmall}
\end{eqnarray}
where 
\begin{eqnarray}
w \equiv &&   \phi^2_{n_E}(x)  
\label{wclosest}
\end{eqnarray}
is the weight of the eigenfunction of the nearest eigenfunction $\phi_{n_E}$ at site $x$.
 Using the probability distribution $\Pi_N(w)$ discussed in the previous section,
one obtains by this change of variables 
that the probability distribution of $G^R_{xx})$
displays the following power-law tail 
\begin{eqnarray}
P(G^R_{xx})  && \opsimeq_{\vert G^R_{xx} \vert \to +\infty} \int_0^{1} dw \Pi_N(w) \int_0^{+\infty} ds R(s)
\delta\left( \vert G^R_{xx} \vert - \frac{ N \rho(E) w  }{  s  } \right)
\nonumber \\
&& \opsimeq_{\vert G^R_{xx} \vert \to +\infty} \int_0^{1} dw \Pi_N(w) \int_0^{+\infty} ds R(s)
\frac{ \delta\left( s - \frac{ N \rho(E) w  }{  \vert G^R_{xx} \vert  } \right) }{\frac{ N \rho(E) w  }{  s^2  }} 
\nonumber \\ && \opsimeq_{\vert G^R_{xx} \vert \to +\infty} 
 \frac{N  \rho(E) \int_0^{+\infty} dw \Pi_N(w) w }{(G^R_{xx})^{2}   }
\label{ptailgrcal}
\end{eqnarray}
Taking into account the first moment of the weight distribution (Eq. \ref{wnav}),
one obtains the Eq 13 of \cite{cizeau}
\begin{eqnarray}
P(G^R_{xx})  && \opsimeq_{\vert G^R_{xx} \vert \to +\infty} 
 \frac{  \rho(E) }{(G^R_{xx})^{2}   }
\label{ptailgr}
\end{eqnarray}
The amplitude of the heavy-tail is simply given by the global density of states
$\rho(E)$ and does not contain any information on the localizations properties
of the eigenvectors.

As a final remark, it is interesting to stress that in similar observables where the denominator does not involve
an external energy $E$ but differences $(E_n-E_m)$ between two energies of the spectrum,
the level repulsion exponent $\beta$ enters the analysis in $R(s) \propto s^{\beta}$
and thus appears in the exponent of the heavy-tails : this phenomenon has been much studied in 
the context of the energy-level-curvature in Anderson localization models \cite{edwards_thouless,mcmillan_anderson,kravtsov,montambaux,kramer,evangelou}
in quantum chaotic systems and random matrices \cite{gaspard,delande,vonOppen,fyodorov_sommers_rmt}, and in Many-Body-Localization models 
\cite{eisert,c_mblflux}.

\subsection { Heavy tail analysis for the probability distribution of $F^R_{xx}=-\partial_E G^R_{xx}$  }

As proposed in Ref. \cite{cizeau}, it is also interesting to consider the derivative with respect to the energy
\begin{eqnarray}
F^R_{xx}(E) && = - \partial_E G_{xx}^R(E) = \sum_{n=1}^N \frac{ \phi^2_n(x)  } { (E-E_n)^2 }  
\label{fderi}
\end{eqnarray}
because, as a consequence of the analytic properties of the Green function, it corresponds also to
the derivative of the imaginary part of the Green function $ G^I_{xx}(z=E-i \eta)$ (Eq \ref{grgi}) with respect to $\eta$ taken at $\eta \to 0$
\begin{eqnarray}
F^R_{xx}(E) && = \left[ \frac{ \partial G^I_{xx}(E-i \eta) }{ \partial \eta } \right]_{\eta=0}
\label{fderianaly}
\end{eqnarray}
Further explanations on the link with the localization criterion based on the
imaginary part of the self-energy can be found in Appendix C of Ref  \cite{cizeau}.

As above, it can become arbitrarily large $F^R_{xx} \to +\infty$
only if the external energy $E$ becomes very close $s \to 0$ to the 
nearest energy-level $E_{n_E}$ in Eq. \ref{diffe}, and it is then given by
\begin{eqnarray}
F^R_{xx} &&  \opsimeq_{ s \to 0} \frac{ \phi^2_{n_E}(x)  }{ \Delta^2 s^2  } = \frac{ N^2 \rho^2(E) w  }{  s^2  } 
\label{fr}
\end{eqnarray}
This change of variables yields the heavy tails (Eq 24 of Ref \cite{cizeau})
\begin{eqnarray}
P(F^R)  && \opsimeq_{ F^R_{xx}   \to +\infty} \int_0^{+\infty} dw \Pi_N(w) \int_0^{+\infty} ds R(s)
\delta\left( F^R_{xx}  -\frac{  N^2 \rho^2(E) w } { s^2  } \right)
\nonumber \\
&& \opsimeq_{F^R  \to +\infty} \int_0^{+\infty} dw \Pi_N(w) \int_0^{+\infty} ds R(s)
\frac{ \delta\left( s- N \rho(E) \sqrt{\frac{ w } {   F^R   }} \right) }
{ 2 N^2 \rho^2(E) \frac{w}{  s^3} }
\nonumber \\
 && \opsimeq_{ F^R_{xx}   \to +\infty} 
 \frac{  \rho(E) N \int_0^{+\infty} dw \Pi_N(w) w^{\frac{1}{2}} }{(F^R_{xx} )^{\frac{3}{2}}   }
= \frac{  \rho(E) Y_{q=\frac{1}{2}}(N) }{(F^R_{xx} )^{\frac{3}{2}}   }
\label{ptailfr}
\end{eqnarray}
So here in contrast to Eq. \ref{ptailgr},
the amplitude of the heavy-tail contains some information on the localization properties
via the generalized moment $Y_{q=\frac{1}{2}}(N) $ of Eq. \ref{yqtauq}
for the index $q=1/2$.

\subsection { Heavy tail analysis for the probability distribution of $G^R_{x \ne y}$  }

As above, the Green function involving two different sites $x \ne y$
\begin{eqnarray}
G^R_{xy}(E) && =\sum_{n=1}^N 
\frac{ \phi_n(x) \phi_n(y)  } { E-E_n }  
\label{grxyeta0diff}
\end{eqnarray}
can become arbitrarilty large $\vert G^R_{xy} \vert \to +\infty$
only if the external energy $E$ becomes very close $s \to 0$ to the 
nearest energy-level $E_{n_E}$ in Eq. \ref{diffe}, and it is then given by
\begin{eqnarray}
\vert G^R_{xy}(E)  \vert &&   \opsimeq_{ s \to 0}  \frac{ \sqrt{ w_1 w_2 } }{ \Delta s  } = \frac{ N \rho(E) \sqrt{ w_1 w_2 }  }{  s  } 
\label{grxyssmall}
\end{eqnarray}
where 
\begin{eqnarray}
w_1 \equiv &&   \phi^2_{n_E}(x)  
\nonumber \\
w_2 \equiv &&   \phi^2_{n_E}(y)  
\label{w1w2}
\end{eqnarray}
are the weights of the eigenfunction $\phi_{n_E}$ on the two sites $x \ne y$.
If one assumes that these two weights are independently drawn with the probability distribution $\Pi_N(w)$,
one obtains the following tails
\begin{eqnarray}
P(G^R_{xy})  && \opsimeq_{\vert G^R_{xy} \vert \to +\infty} \int_0^{1} dw_1 \Pi_N(w_1) int_0^{1} dw_2 \Pi_N(w_2) \int_0^{+\infty} ds R(s)
\delta\left( \vert G^R_{xy} \vert - \frac{ N \rho(E) \sqrt{ w_1 w_2 }    }{  s  } \right)
\nonumber \\
&& \opsimeq_{\vert G^R_{xy} \vert \to +\infty}  \int_0^{1} dw_1 \Pi_N(w_1) int_0^{1} dw_2 \Pi_N(w_2) \int_0^{+\infty} ds R(s)
\frac{ \delta\left( s - \frac{ N \rho(E) \sqrt{ w_1 w_2 }    }{  \vert G^R_{xx} \vert  } \right) }{\frac{ N \rho(E) \sqrt{ w_1 w_2 }    }{  s^2  }} 
\nonumber \\ && \opsimeq_{\vert G^R_{xy} \vert \to +\infty} 
 \frac{N  \rho(E) [ \int_0^{+\infty} dw \Pi_N(w) \sqrt{ w } ]^2  }{(G^R_{xy})^{2}   }
=  \frac{  \rho(E) \left( \frac{[Y_{q=\frac{1}{2}} (N)]^2}{N} \right)  }{(G^R_{xy})^{2}   }
\label{ptailgrcalxy}
\end{eqnarray}
In particular, the scaling with respect to the size $N$ of the amplitude of this heavy-tail
\begin{eqnarray}
\frac{[Y_{q=\frac{1}{2}} (N)]^2}{N} \propto N^{-2 \Delta(q=\frac{1}{2}) }
\label{amplixyqdemi}
\end{eqnarray}
involves the anomalous dimension $\Delta(q)$ of Eq. \ref{Deltaqtauq}
for the index $q=\frac{1}{2}$.

\subsection{ Discussion}

For $\eta=0$, we have described how
various heavy-tails appear in the probability distribution of the Green function or of its derivative with respect
to the energy : while the amplitude of the tail in Eq. \ref{ptailgr} only contains the global density of states,
the amplitudes of the heavy-tails of Eq. \ref{ptailfr} and of Eq \ref{ptailgrcalxy}
both involve the generalized moment $Y_{q=\frac{1}{2}}$ that includes some information on the localization 
properties of the eigenvectors, although it is limited to the specific value $q=\frac{1}{2}$.

In the remaining sections of this paper, we will 
see how the introduction of some finite-size imaginary part $\eta=\frac{c}{N^{\delta}}$
  leads to much richer statistical properties for the imaginary part of the Green function,
that will involve all the generalized moments $Y_q(N)$ of arbitrary index $q$.

\section{ Imaginary part of the Green function as a function of the broadening $\eta(N)$ }

\label{sec_imaginary}

In the remaining of the paper, we focus
on the imaginary part of the Green function at coinciding points $x=y$
\begin{eqnarray}
 G^I && =   \sum_{n=1}^N w_n \frac{  \eta} { (E-E_n)^2 + \eta^2 }  
\label{gimaginary}
\end{eqnarray}
invoving the weights
\begin{eqnarray}
w_n \equiv \phi_n^2(x) 
\label{wnx}
\end{eqnarray}

\subsection{ Averaged Value   }

 Since the first moment of the weight $w=\phi_n^2(x) $ is fixed by Eq. \ref{wnav},
the averaged value of Eq. \ref{gimaginary} 
\begin{eqnarray}
\overline{  G^I } && =   \sum_{n=1}^N \ \ \ \overline{  w_n } \ \ \   \overline{  \frac{  \eta} { (E-E_n)^2 + \eta^2 }  }
\nonumber \\ 
&& = N \frac{1}{N} \int_{-\infty}^{+\infty} dE_n \rho(E_n) \frac{  \eta} { (E-E_n)^2 + \eta^2 }  
\nonumber \\ 
&& =     \int_{-\infty}^{+\infty}  du \rho(E+ \eta u) \frac{ 1} { u^2 + 1} 
\nonumber \\
&& \opsimeq_{\eta \to 0} \pi \rho(E)  
\label{giav}
\end{eqnarray}
does not contain any information on the localization properties,
and does not depend on the scaling relation between $\eta$ et $N$.

\subsection{ Square regularization of the delta function   }

Since the Lorentzian is some regularization of the delta function with some width $\eta$,
it is convenient to replace it by the simpler square regularization of width $\eta$
\begin{eqnarray}
 \frac{  \eta} { (E-E_n)^2 + \eta^2 }  \opsimeq_{\eta \to 0} \pi \delta(E-E_n)
\opsimeq_{\eta \to 0}   \frac{\pi }{ \eta} \theta\left( - \frac{\eta}{2} \leq E-E_n \leq \frac{\eta}{2}  \right)
\label{reguladelta}
\end{eqnarray}
Then the imaginary part $G^I$ of the Green function becomes a sum over the weights $w_n$ 
(Eq. \ref{wnx}) for the levels in the energy-window $E - \frac{\eta}{2} \leq E_n \leq E+\frac{\eta}{2}$
\begin{eqnarray}
 G^I && \simeq    \frac{\pi }{ \eta} \sum_{n=1}^N w_n \theta\left( E - \frac{\eta}{2} \leq E_n \leq E+\frac{\eta}{2}  \right)
\label{gsquare}
\end{eqnarray}
It is then clear that the statistics will strongly depend on the number of states in this energy-window
\begin{eqnarray}
 M \equiv    \sum_{n=1}^N  \theta\left( E - \frac{\eta}{2} \leq E_n \leq E+\frac{\eta}{2}  \right)
\label{mrandom}
\end{eqnarray}

\subsection{ Role of the scaling of the broadening $\eta=\frac{c}{N^{\delta}}$ }

The average value of Eq. \ref{mrandom}
is given by the ratio between the broadening $\eta=\frac{c}{N^{\delta}}$ and the level spacing $\Delta=\frac{1}{N \rho(E)}$
\begin{eqnarray}
 M_{av} =    \eta N \rho(E) = \frac{\eta}{\Delta} = c \rho(E) N^{1-\delta} 
\label{mav}
\end{eqnarray}
so that one has to distinguish the following cases :

(i) The standard scaling $\delta=1$ correspond to a fixed number of states in the energy-window
\begin{eqnarray}
 M_{av} =    c \rho(E) 
\label{mav1}
\end{eqnarray}
The sum of a finite number of multifractal weights is then expected to produce strong fluctuations for the imaginary part of Eq. \ref{gsquare}.
In particular, the typical value of the imaginary Green function is expected to involve the typical exponent $\alpha_0$ (Eq. \ref{alphazero})
\begin{eqnarray}
 G^I_{typ}&& \propto N^{1-\alpha_0}
\label{gityp}
\end{eqnarray}
The decay with the size $N$ is
thus completely different from the averaged value of Eq. \ref{giav}
that does not decay with $N$.

(ii) The scaling $\delta=0$ where $\eta=c \ll 1$ is small but independent on the size $N$
corresponds to an extensive number of states in the energy-window
\begin{eqnarray}
 M_{av} =    c \rho(E) N 
\label{mav3}
\end{eqnarray}
Then one expects that this sum over an extensive number of weights will be effective to reproduce the average value of Eq. \ref{wnav}
so that the imaginary Green function remains concentrated around
 its averaged value
\begin{eqnarray}
 G^I && \simeq    \frac{\pi }{ \eta} \frac{M}{N} = \pi \rho(E)
\label{gaveff}
\end{eqnarray}

(iii) The non-standard scaling $0<\delta<1$ mentioned in the Introduction corresponds to a sub-extensive growth of the number of states
\begin{eqnarray}
 M_{av} =     c \rho(E) N^{1-\delta} 
\label{mav2}
\end{eqnarray}
 To characterize the statistics of the imaginary Green function 
in this regime, one needs to analyze the statistics of 
a sub-extensive number $M_{av}$ of  multifractal weights.

In the following sections, we analyze in details the regime
(i) of the standart scaling $\delta=1$
and and the regime (iii) of the non-standard scaling $0<\delta<1$.

\section{ Statistics of the imaginary Green function for $ \eta=\frac{c}{N} $ with $c \ll 1$  }

\label{sec_onelevel}

In this section, we consider the scaling $ \eta=\frac{c}{N}  $ with $c \ll 1$ : 
this corresponds to the regime where the averaged number of states in the energy-window
is independent of $N$ but very small (Eq. \ref{mav1})
\begin{eqnarray}
 M_{av} =    c \rho(E) \ll 1
\label{mav1csmall}
\end{eqnarray}

Then one needs to consider only the nearest level
 $\vert E_{n_E}-E \vert =s \Delta $ as in Eq. \ref{diffe},
where $\Delta=\frac{1}{N \rho(E)}$ is the level spacing and 
where $s$ is the $O(1)$ random variable distributed exponentially $R(s)=e^{-s}$ (Eq \ref{rsexp})
and one obtains the approximation
\begin{eqnarray}
G^I \simeq w_{n_E}  \frac{  \eta} { \Delta^2 s^2+\eta^2 } =   w_{n_E}  \frac{ c N \rho^2(E) } {  s^2+c^2 \rho^2(E)  }
\label{ginearest}
\end{eqnarray}

\subsection{ Scaling of the moments }

Eq. \ref{ginearest} yields that the moment of order $q$ reads
\begin{eqnarray}
\overline{ (G^I)^q } \simeq \overline{ (w_{n_E} )^q }   \int_0^{+\infty} ds e^{-s} \left(  \frac{ c N \rho^2(E) } { s^2+c^2 \rho^2(E)  }  \right)^q
= \frac{Y_q(N)}{N }  \int_0^{+\infty} ds e^{-s} \left(  \frac{ c N \rho^2(E) } { s^2+c^2 \rho^2(E) }  \right)^q
\label{giqnearest}
\end{eqnarray}
that one needs to evaluate for $c \ll 1$.  One has to distinguish two regions :

(a) In the region $ q<\frac{1}{2}$ where $s^{-2q}$ is integrable near the origin $s \to 0$, one obtains the leading behavior
\begin{eqnarray}
\left[ \overline{ (G^I)^q } \right]_{ q < \frac{1}{2}} 
&& \opsimeq_{c \to 0}    \frac{Y_q(N)}{N }  \int_0^{+\infty} ds e^{-s}  \frac{ [c N \rho^2(E)]^q } { s^{2q} }  
\nonumber \\
&&   \opsimeq_{c \to 0}   Y_q(N) N^{q-1}   c^q  \rho^{2q}(E) \Gamma(1-2q)
\label{giqnearestsmall}
\end{eqnarray}

(b) In the region $q>\frac{1}{2}$, where $s^{-2q}$ is not integrable near the origin $s \to 0$,
one needs to make the change of variable $s= c \rho(E) t$ in Eq. \ref{giqnearest}
to obtain the leading behavior 
\begin{eqnarray}
\left[ \overline{ (G^I)^q }  \right]_{ q > \frac{1}{2}} && \simeq  
\frac{Y_q(N)}{N }  \int_0^{+\infty} c \rho(E) dt e^{- c \rho(E) t} \left(  \frac{  N  } { (t^2+1) c }  \right)^q
\nonumber \\
&& \simeq Y_q(N) N^{q-1} c^{1-q} \rho(E)  \int_0^{+\infty}   dt e^{- c \rho(E) t}  \frac{  1  } { (t^2+1)^q  }  
\nonumber \\
&&  \opsimeq_{c \to 0}   Y_q(N) N^{q-1} c^{1-q} \rho(E)  \int_0^{+\infty}   dt  \frac{  1  } { (t^2+1)^q  }  
\nonumber \\
&&   \simeq     Y_q(N) N^{q-1} c^{1-q} \rho(E)    \frac{ \sqrt{\pi} \Gamma \left(q-\frac{1}{2} \right) }{\Gamma (q) }
\label{giqnearestbigm}
\end{eqnarray}
that reproduces in particular the exact averaged value $\overline{ G^I } = \pi \rho(E)$ of Eq. \ref{giav} for $q=1$.

Using the multifractal anomalous exponent of Eq. \ref{Deltaqtauq}
\begin{eqnarray}
Y_q(N) \propto N^{1-q-\Delta(q)}
\label{yqanomalous}
\end{eqnarray}
on obtains that for both regions $q<\frac{1}{2}$ and $q>\frac{1}{2}$, the scaling with respect to the size $N$ is given by
\begin{eqnarray}
\overline{ (G^I)^q }  &&    \propto  Y_q(N) N^{q-1} \propto N^{-\Delta(q)}
\label{giqnearestmultif}
\end{eqnarray}
in agreement with the usual expectation of Eq. \ref{ldos1statedeltaq} for the local density of states.

\subsection{ Probability distribution  }

Besides the moments discussed above, it is interesting to
 consider the probability distribution 
produced by the change of variables of Eq. \ref{ginearest}
\begin{eqnarray}
P(G^I)  && \simeq
 \int_0^{1} dw \Pi_N(w) \int_0^{+\infty} ds e^{-s} 
\delta\left(  G_I  -  w_{n_E}  \frac{ c N \rho^2(E) } {  s^2+c^2 \rho^2(E)  } \right)
\nonumber \\
&& \simeq
 \int_0^{1} dw \Pi_N(w) \int_0^{+\infty} ds e^{-s}
\theta \left( c G^I  \leq w N  \right)
 \delta\left(  s  - \rho(E) \sqrt{ \frac{w N c}{G^I} \left(1- \frac{c G^I}{w N} \right) }  \right) 
\frac{ \rho(E) \sqrt{cN} \sqrt{w} }{ (G^I)^{\frac{3}{2}}  2 \sqrt{1-\frac{c G^I}{w N} }}
\nonumber \\
&& \simeq \frac{ \rho(E) \sqrt{cN}  }{ 2 (G^I)^{\frac{3}{2}}   }
 \int_0^{1} dw \Pi_N(w)  e^{- \rho(E) \sqrt{ \frac{w N c}{G^I} \left(1- \frac{c G^I}{w N} \right) }}
\theta \left( c G^I  \leq w N  \right)
\frac{  \sqrt{w} }{  \sqrt{1-\frac{c G^I}{w N} }}
\label{tailgi}
\end{eqnarray}
The power-law factor $1/(G^I)^{\frac{3}{2}}$ explains the change at $q=1/2$
 found above for the moments (Eqs \ref{giqnearestsmall} and \ref{giqnearestbigm}) :
in particular, all the moments for $q>\frac{1}{2}$ are actually dominated by the cut-off at large scale $\theta \left(  G^I  \leq \frac{w N }{c} \right)$.

The presence of the power-law factor $1/(G^I)^{\frac{3}{2}}$ in the probability distribution of the imaginary Green function
has been obtained previously for various specific models,
in particular for the L\'evy matrix model (Eq C3a of Ref \cite{cizeau}),
for the Generalized Rosenzweig-Potter model (Eq 9 of Ref \cite{biroli_rosen}),
and for Anderson Localization on the Bethe Lattice (Eq 43 of Ref \cite{mirlin_tik}).

\section{ Statistics of the imaginary Green function for $ \eta=\frac{c}{N} $ with $c \gg 1$  }

\label{sec_many}

In this section, we consider the scaling $ \eta=\frac{c}{N}  $ with $c \gg 1$ : 
this corresponds to the regime where the averaged number of states in the energy-window
is independent of $N$ but very large (Eq. \ref{mav1})
\begin{eqnarray}
 M_{av} =    c \rho(E) \gg 1
\label{mav1cbig}
\end{eqnarray}
so that the imaginary Green function of Eq. \ref{gsquare} involves a sum over a large number $M$ of weights $w_i=N^{-\alpha_i}$
\begin{eqnarray}
 G^I && \simeq    \frac{\pi  }{ \eta } \sum_{i=1}^M w_i =  \frac{\pi N }{ c } \sum_{i=1}^M N^{-\alpha_i}
\label{gsquaremi}
\end{eqnarray}

\subsection{ Role of the minimal value $\alpha_{min}$ among $M$ exponents }

Let us introduce the minimal exponent $\alpha_{min} $ and the maximal exponent
$\alpha_{max} $  among these $M$ random exponents $\alpha_i$
drawn independently with the probability distribution $\pi_N(\alpha) \propto N^{f(\alpha)-1}$ (Eq. \ref{pialpha}).
The discrete sum of Eq. \ref{gsquaremi} over a large number $M$ of terms
can be then approximated by the integral
\begin{eqnarray}
 G^I &&  \simeq   \frac{\pi N }{ c } M  \int_{\alpha_{min}}^{\alpha_{max}} d \alpha N^{f(\alpha)-1} N^{-\alpha}
\nonumber \\
&& \simeq   \pi \rho(E)   \int_{\alpha_{min}}^{\alpha_{max}} d \alpha N^{f(\alpha)-\alpha}
\label{gsquarem}
\end{eqnarray}
This integral over $\alpha$ is exactly the same as for the generalized moment $Y_{q=1}(N)$ of Eq \ref{yqsaddle}
except for the restricted domain of integration $\alpha_{min}<\alpha<\alpha_{max}$,
so that there are two possibilities, depending on the position of the saddle-point value $\alpha_{q=1} = f(\alpha_1) = D(1)$ of Eq. \ref{alphaun} :

(a) If $\alpha_{min} \leq \alpha_{q=1} $ then the saddle-point calculus of Eq. \ref{yqsaddle}
can be applied to obtain with $\tau(q=1)=0$
\begin{eqnarray}
G^I &&  \simeq    \pi  \rho(E)  N^{-\tau(q=1)} = \pi  \rho(E) 
\label{gversav}
\end{eqnarray}
corresponding to the averaged value of Eq \ref{giav}.

(b) If $ \alpha_{q=1} < \alpha_{min}$ then the saddle-point calculus in Eq. \ref{gsquarem} takes place at the boundary $\alpha_{min}$
and leads to
\begin{eqnarray}
 G^I &&  \simeq    \pi \rho(E)  N^{f(\alpha_{min})-\alpha_{min}} 
\label{gsdomini}
\end{eqnarray}
In this region, the statistics of $G^I$ is thus directly related to the statistics of $\alpha_{min}$.
The cumulative distribution of $  {\rm min}(\alpha_1,...,\alpha_M) $ reads for large $M$ (see for instance the books on extreme value statistics \cite{extreme})
\begin{eqnarray}
Prob( \alpha_{min} \leq {\rm min}(\alpha_1,...,\alpha_M)  ) && = \prod_{i=1}^M \int_{\alpha_{min}}^{+\infty} d \alpha_i \pi_N(\alpha_i) 
  =  \left[  1- \int_{0}^{\alpha_{min}} d \alpha \pi_N(\alpha) \right]^M
\nonumber \\
&&  \opsimeq _{M \gg 1} e^{- M   \int_{0}^{\alpha_{min}} d \alpha  \pi_N(\alpha)  }
\label{cumulalphamin}
\end{eqnarray}

\subsection{ Probability distribution of $\alpha_{min}$ in the typical region }

As usual in extreme value statistics \cite{extreme},
it is convenient to introduce the characteristic scale $a_M$ of $\alpha_{min}$ 
corresponding to a fixed value $e^{-1}$ in the cumulative distribution of Eq. \ref{cumulalphamin}
\begin{eqnarray}
 1 = M   \int_{0}^{a_M} d \alpha  \pi_N(\alpha)  
\label{defam}
\end{eqnarray}
Even if $M$ is large (Eq \ref{mav1cbig}), it does not grow with $N$,
so that $a_M$ remains in the vicinity of the typical value $\alpha_0$, where the multifractal spectrum can generically
be expanded into the quadratic form
\begin{eqnarray}
f(\alpha) = 1 - \frac{(\alpha-\alpha_0)^2 }{2 \sigma^2} +o((\alpha-\alpha_0)^2)
\label{fgaussexp}
\end{eqnarray}
so that the probability distribution $\pi_N(\alpha) $ can be approximated by the Gaussian distribution in this region
\begin{eqnarray}
\pi_N(\alpha) \simeq \sqrt{ \frac{  \ln N }{ 2 \pi \sigma^2   } } e^{- \frac{ (\alpha -\alpha_0 )^2 }{2  \sigma^2  } \ln N }
\label{pialphagauss}
\end{eqnarray}
To evaluate Eq. \ref{cumulalphamin}, we need the integral
\begin{eqnarray}
\int_{0}^{\alpha_{min}}  d \alpha \pi_N(\alpha) && = \int_{0}^{\alpha_{min}}
  d \alpha\sqrt{ \frac{  \ln N }{ 2 \pi \sigma^2   } } e^{- \frac{ (\alpha -\alpha_0 )^2 }{2  \sigma^2  } \ln N}
\nonumber \\
&& =  \sqrt{ \frac{  \ln N }{ 2 \pi \sigma^2   } } e^{- \frac{ (\alpha_{min} -\alpha_0 )^2 }{2  \sigma^2  } \ln N}
\int_0^{\alpha_{min} }  dv
 e^{- \frac{ v (\alpha_0 - \alpha_{min}) }{ \sigma^2  } \ln N - \frac{v^2 }{2  \sigma^2  } \ln N }
\nonumber \\
&& =  \sqrt{ \frac{  \ln N }{ 2 \pi \sigma^2   } } N^{- \frac{ (\alpha_{min} -\alpha_0 )^2 }{2  \sigma^2  } }
\int_0^{\frac{v (\alpha_0-\alpha_{min}) \ln N}{\sigma^2} }  dy \frac{\sigma^2 }{(\alpha_0 - \alpha_{min}) \ln N } 
 e^{- y -   \frac{\sigma^2 y^2 }{ 2 (\alpha_0 - \alpha_{min})^2 (\ln N) }  }
\nonumber \\
&& \opsimeq_{N \gg 1}  
 \sqrt{ \frac{  \sigma^2}{ 2 \pi \ln N (\alpha_0 - \alpha_{min})^2 } }
 N^{- \frac{ (\alpha_{min} -\alpha_0 )^2 }{2  \sigma^2  }} 
\label{piacumul}
\end{eqnarray}

The characteristic scale $a_M$ defined by Eq. \ref{defam}
\begin{eqnarray}
1 && =  M   \int_{-\infty}^{a_M} d \alpha \pi_N(\alpha) 
 \simeq M   \sqrt{ \frac{  \sigma^2}{ 2 \pi \ln N (\alpha_0 - a_M)^2 } }
 N^{- \frac{ (a_M -\alpha_0 )^2 }{2  \sigma^2  }} 
\label{amcalcul}
\end{eqnarray}
is thus given at leading order by
\begin{eqnarray}
 a_M    \simeq  \alpha_0- \sqrt{ 2  \sigma^2 \frac{ \ln M -\frac{1}{2}  \ln \left(  4 \pi  \ln M   \right) }{\ln N } }
\label{amres}
\end{eqnarray}
In particular since $M$ remains fixed as $N$ grows, it remains close to the typical value $\alpha_0$
so that the expansion of Eq. \ref{fgaussexp} is consistent.

As usual in extreme value statistics \cite{extreme},
we now make the change of variables
\begin{eqnarray}
 \alpha_{min}=a_M+b_M \xi
\label{bmxi}
\end{eqnarray}
to look for the scale $b_M$ that will lead to
 a rescaled random variable $\xi$ of order $O(1)$.
The cumulative distribution of Eq. \ref{cumulalphamin} becomes
\begin{eqnarray}
Prob( \alpha_{min}=a_M+b_M \xi  \leq min(\alpha_1,...,\alpha_M)  ) && 
\simeq e^{- M   \int_{-\infty}^{a_M+b_N \xi} d \alpha  \pi_N(\alpha)  }
\nonumber \\
&& \simeq e^{- M    
\sqrt{ \frac{  \sigma^2}{ 2 \pi \ln N (\alpha_0 - a_M)^2 } }
 N^{- \frac{ (a_M -\alpha_0 )^2 }{2  \sigma^2  }}
 N^{ \frac{  (\alpha_0-a_M )b_M \xi }{  \sigma^2  }}
 N^{- \frac{ b_N^2 \xi^2   }{2  \sigma^2 }}}
\nonumber \\
&& \simeq e^{-  e^{ \xi  \frac{ b_M  (\alpha_0-a_M ) \ln N }{  \sigma^2  }}
 e^{- \frac{ b_N^2 \xi^2   }{2  \sigma^2 }} \ln N }
\label{cumulcalcul}
\end{eqnarray}
This yields that the appropriate scale $b_M$ for the width reads
\begin{eqnarray}
b_M   = \frac{ \sigma^2 }{(\alpha_0-a_M ) \ln N} 
\simeq   \sqrt{   \frac{\sigma^2    }
{2 \ln N (  \ln M -\frac{1}{2}  \ln \left(  4 \pi  \ln M   \right) ) } }
\label{bmres}
\end{eqnarray}
and then the distribution of the rescaled variable $\xi$ is the 
well-known Gumbel distribution \cite{extreme}
\begin{eqnarray}
Prob( \alpha_{min}=a_M+b_N \xi  \leq min(\alpha_1,...,\alpha_M)  ) &&   \simeq e^{-  e^\xi }
\label{cvgumbel}
\end{eqnarray}
The probability distribution obtained by derivation of the cumulative distribution reads
\begin{eqnarray}
 g(\xi)  = - \frac{d}{d\xi} (e^{-  e^\xi } ) =e^{ \xi -  e^\xi }
\label{gumbel}
\end{eqnarray}

\subsection{ Probability distribution of $G^I$ in the typical region }

In this region, the imaginary part of the Green function (Eq. \ref{gsquarem}) is given by
\begin{eqnarray}
 G^I && \simeq   \pi \rho(E)  N \int_{\alpha_{min}}^{\alpha_{max}} d \alpha \pi_N(\alpha) N^{-\alpha}
\nonumber \\
&& \simeq  \pi \rho(E)  N \int_{\alpha_{min}}^{\alpha_{max}} d \alpha \sqrt{ \frac{  \ln N }{ 2 \pi \sigma^2   } } e^{- \frac{ (\alpha -\alpha_0 )^2 }{2  \sigma^2  } \ln N } e^{-\alpha \ln N }
\nonumber \\
&& \simeq   \pi  \rho(E) N 
\sqrt{ \frac{  \ln N }{ 2 \pi \sigma^2   } }
 N^{- \frac{ (\alpha_{min} -\alpha_0 )^2 }{2  \sigma^2  } }
 N^{-\alpha_{min}}
\int_{0}^{\alpha_{max}-\alpha_{min}} dv e^{-v \ln N }
 e^{ \frac{  v (\alpha_0 - \alpha_{min}  ) }{  \sigma^2  } \ln N}
 e^{- \frac{ v^2 }{2  \sigma^2  } \ln N } 
\nonumber \\
&& \simeq   \pi  \rho(E) N 
\sqrt{ \frac{ 1 }{ 2 \pi \sigma^2 \ln N  } }
 N^{- \frac{ (\alpha_{min} -\alpha_0 )^2 }{2  \sigma^2  } }
 N^{-\alpha_{min}}
\label{gigauss}
\end{eqnarray}

The replacement $\alpha_{min}=a_M+b_M \xi $ of Eq. \ref{bmxi} yields
\begin{eqnarray}
 G^I && \simeq    \pi  \rho(E) N 
\sqrt{ \frac{ 1 }{ 2 \pi \sigma^2 \ln N  } }
 N^{- \frac{ (a_M -\alpha_0 )^2 }{2  \sigma^2  } }   N^{-a_M}
 N^{ \left[ \frac{ (\alpha_0-a_M ) }{  \sigma^2  }  -1 \right] b_M \xi }   
 N^{- \frac{ b_M^2 \xi^2  }{2  \sigma^2  } }
\label{gigaussxi}
\end{eqnarray}
Since Eq \ref{amres} yields $\frac{ (\alpha_0-a_M ) }{  \sigma^2  } \simeq \sqrt{ 2   \frac{ \ln M  }{ \sigma^2\ln N } } \ll 1$
and Eq. \ref{bmres} yields $
b_M^2 \ln N   \simeq 
      \frac{\sigma^2    }
{2  (  \ln M -\frac{1}{2}  \ln \left(  4 \pi  \ln M   \right) ) } \ll 1$,
 Eq. \ref{gigaussxi} simplifies into
\begin{eqnarray}
 G^I && \simeq   G^I_{typ}
 e^{ -  \xi b_M \ln N }   
\label{gxi}
\end{eqnarray}
where $\xi$ is distributed with the Gumbel distribution of Eq. \ref{gumbel},
and where the typical value defined as the value for $\xi=0$ reads using Eq \ref{amres}
\begin{eqnarray}
 G^I_{typ} && \simeq \pi  \rho(E) N 
\sqrt{ \frac{ 1 }{ 2 \pi \sigma^2 \ln N  } }
 N^{- \frac{ (a_M -\alpha_0 )^2 }{2  \sigma^2  } }   N^{-a_M}
\nonumber \\
&& \simeq   \pi  \rho(E) \frac{N}{M} 
\sqrt{ \frac{ 2 \ln M }{  \sigma^2 \ln N  } } N^{-a_M}
\nonumber \\
&& \simeq  \pi  \rho(E) \frac{N^{1- \alpha_0 } }{M} 
\sqrt{ \frac{ 2 \ln M }{  \sigma^2 \ln N  } } 
 e^{ \sqrt{ 2  \sigma^2 \ln N [ \ln M -\frac{1}{2}  \ln \left(  4 \pi  \ln M   \right) ]}}
\label{gtypgauss}
\end{eqnarray}
In particular, even if there are logarithmic corrections, 
the leading behavior in $N$ is nevertheless
 given by $N^{1-\alpha_0}$ and involves the typical exponent $\alpha_0$.

Using Eq. \ref{bmres}, i is now useful to introduce the small parameter 
\begin{eqnarray}
\mu_N \equiv  \frac{1}{ b_M \ln N} =   \sqrt{   \frac
{2  (  \ln M -\frac{1}{2}  \ln \left(  4 \pi  \ln M   \right) ) }{\sigma^2     \ln N  } }  
\opsimeq \sqrt{   \frac{2    \ln M}{\sigma^2  \ln N   } }
\label{musmall}
\end{eqnarray}

Then the change of variables between the Gumbel distributed variable $\xi$ (Eq. \ref{gumbel}) and the imaginary Green function of Eq. \ref{gxi}
\begin{eqnarray}
 G^I && \simeq   G^I_{typ}  e^{ - \frac{ \xi }{\mu_N}  }   
\label{gximu}
\end{eqnarray}
yields that the probability distribution of $G^I$ in the scaling region 
\begin{eqnarray}
P( G^I ) && = \int_{-\infty}^{+\infty} d\xi g(\xi) \delta\left( G^I - G^I_{typ}
  e^{- \frac{\xi}{\mu_N} } \right)
\nonumber \\
&& = \frac{ \mu_N (G^I_{typ} )^{\mu_N} }{ (G^I)^{1+\mu_N} } e^{- \left( \frac{G^I_{typ}}{G^I} \right)^{\mu_N} }
\label{frechet}
\end{eqnarray}
is a Fr\'echet distribution \cite{extreme}
with a vanishing exponent $\mu_N$ (Eq. \ref{musmall}).
In particular, this means that all moments $\overline{(G^I)^q} $ of index $q>0$ are actually governed by
 rare events outside this scaling region, and should be thus evaluated 
from the statistics of $\alpha_{min}$ outside the scaling region of Eq. \ref{cvgumbel}
as we now describe.

\subsection{ Multifractal properties of $\alpha_{min}$ and $G^I$ }

To evaluate the probability of rare events where $G^I$ is anomalously large with respect to the typical value,
we need to evaluate the probability of rare events where $\alpha_{min}$ is anomalously small $\alpha_{min} \ll a_M$, i.e. 
the cumulative probability of Eq. \ref{cumulalphamin}
\begin{eqnarray}
Prob( \alpha_{min} \leq {\rm min}(\alpha_1,...,\alpha_M)  ) &&  \simeq e^{- M   \int_{0}^{\alpha_{min}} d \alpha  \pi_N(\alpha)  }
\label{cumulalphaminbis}
\end{eqnarray}
is also very small $ M   \int_{0}^{\alpha_{min}} d \alpha  \pi_N(\alpha) \ll 1$.
As a consequence,
the probability distribution obtained by derivation becomes
\begin{eqnarray}
P_{min}( \alpha_{min}  ) 
&& = - \frac{d}{d \alpha_{min} } e^{- M   \int_{-\infty}^{\alpha_{min}} d \alpha  \pi_N(\alpha)  }
= M  \pi_N(\alpha_{min} )e^{- M   \int_{-\infty}^{\alpha_{min}} d \alpha  \pi_N(\alpha)  }
\nonumber \\
&& \opsimeq_{\alpha_{min} \ll a_M} M  \pi_N(\alpha_{min} ) \simeq M N^{f(\alpha_{min})-1}
\label{minlargedev}
\end{eqnarray}
and thus directly reflects the multifractal spectrum $f(\alpha)$.
The physical meaning of Eq. \ref{minlargedev} is that 
once an anomalously small exponent has been drawn,
the constraint that the other $(M-1)$ exponents have to be bigger
disappears because it is satisfied automatically.

When an anomalously small exponent $\alpha_{min}$ has been drawn,
 the imaginary Green function of Eq. \ref{gsquaremi}
is dominated by the corresponding anomalously big contribution
\begin{eqnarray}
 G^I && \opsimeq_{G^I \gg G^I_{typ}}  \frac{ \pi }{\eta}  N^{-\alpha_{min}} \simeq \frac{ \pi   }{ c }  N^{1-\alpha_{min}}
\label{gianomalous}
\end{eqnarray}
so that the moments of index can be directly evaluated using Eq. \ref{minlargedev}
\begin{eqnarray}
\overline{ ( G^I)^q } && \simeq   \int d\alpha_{min}  M N^{f(\alpha_{min})-1}
\left[  \frac{ \pi   }{ c }  N^{1-\alpha_{min}} \right]^q
\nonumber \\
&&  \simeq  \frac{ \pi^q   }{ c^q } N^{q-1} M \int d\alpha_{min}   N^{f(\alpha_{min}) -q \alpha_{min}} 
\nonumber \\
&&  \simeq  \rho(E) \pi^q  N^{q-1} c^{1-q}   Y_q(N)
\label{gqano}
\end{eqnarray}
that reproduces in particular the exact averaged value $\overline{ G^I } =\pi \rho(E) $ of Eq. \ref{giav} for $q=1$.

The dependence with respect to the size $N$ thus involves the anomalous exponents $\Delta(q)$ of Eq. \ref{Deltaqtauq}
\begin{eqnarray}
\overline{ ( G^I)^q } && \propto N^{ -\Delta(q) }
\label{gqanoscaling}
\end{eqnarray}
Our conclusion is thus
 that Eq. \ref{ldos1state} is valid
also in the regime $\eta=\frac{c}{N}$ with $c \gg 1$, and does not require 
strong correlation between the weights of consecutive eigenstates :
here we have assumed that the weights were drawn independently with the
multifractal distribution, and we have obtained Eq. \ref{gqanoscaling}
as a consequence of the rare-event analysis described above.

\section{Statistics of the imaginary Green function for  $ \eta = \frac{c}{N^{\delta}} $ with $0<\delta<1$ }

\label{sec_delta}

In this section, we consider the scaling $ \eta \propto \frac{1}{N^{\delta}} $ with $0<\delta<1$ :
this corresponds to the regime where the averaged number of states in the energy-window
grows sub-extensively with $N$ (Eq. \ref{mav2})
\begin{eqnarray}
 M_{av} =    c \rho(E) N^{1-\delta}
\label{mav2sec}
\end{eqnarray}
while the corresponding imaginary Green function is given by (Eq \ref{gsquare})
\begin{eqnarray}
 G^I && \simeq    \frac{\pi N^{\delta} }{ c } \sum_{i=1}^M w_i =  \frac{\pi N^{\delta} }{ c } \sum_{i=1}^M N^{-\alpha_i}
\label{gsquaremd}
\end{eqnarray}

So we have to adapt the analysis of the previous section concerning
the case where $M$ was large but independent of $N$ to the case
where $M$ grows as Eq. \ref{mav2sec}.

\subsection{ Statistical properties in the typical region }

In the typical region, the sum in Eq. \ref{gsquaremd}
can be replaced by an integration over $\alpha$ (as in Eq \ref{gsquarem})
\begin{eqnarray}
 G^I && \simeq    \pi \rho(E) N  \int_{\alpha_{min}}^{\alpha_{max}}   d \alpha  \pi_N(\alpha) N^{-\alpha_i}
\label{gsquaremdd}
\end{eqnarray}
so that one needs to analyze the statistics of $\alpha_{min}$
via its cumulative distribution of Eq. \ref{cumulalphamin} 
\begin{eqnarray}
Prob( \alpha_{min} \leq {\rm min}(\alpha_1,...,\alpha_M)  ) && 
\simeq e^{- M   \int_{0}^{\alpha_{min}} d \alpha  \pi_N(\alpha)  } = e^{-  c \rho(E) N^{1-\delta}    \int_{0}^{\alpha_{min}} d \alpha  \pi_N(\alpha)  } 
\label{cumulalphamind}
\end{eqnarray}

 The characteristic scale $a_M$ of $\alpha_{min}$ can be defined using $\pi_N(\alpha) \propto N^{f(\alpha)-1} $ 
\begin{eqnarray}
 1 =N^{1-\delta}    \int_{0}^{a_M} d \alpha  \pi_N(\alpha) \propto  N^{1-\delta} \pi_N(a_M) \propto   N^{1-\delta}  N^{f(a_M)-1} =  N^{f(a_M)-\delta}
\label{defam2}
\end{eqnarray}
so that at leading order in $N$ it is given by the solution of
\begin{eqnarray}
f(a_M) = \delta 
\label{eqforam}
\end{eqnarray}

As in Eqs \ref{gversav} and \ref{gsdomini}, one has to discuss whether the solution $a_M$ of Eq. \ref{eqforam}
is smaller or bigger than the saddle-point value
 $\alpha_{q=1}=f(\alpha_{q=1} ) =D(1)$ (Eq. \ref{alphaun})
of the generalized moment $Y_{q=1}(N)$  for $q=1$ :

(i) If $a_M \leq \alpha_{q=1} $
 i.e. $f(a_M) =\delta \leq D(1)=f( \alpha_{q=1}) =\alpha_{q=1}  $,
then the saddle-point calculus of Eq. \ref{yqsaddle}
can be applied to obtain with $\tau(q=1)=0$
\begin{eqnarray}
G^I &&  \simeq    \pi  \rho(E)  N^{-\tau(q=1)} = \pi  \rho(E) 
\label{gversavbis}
\end{eqnarray}
corresponding to the averaged value of Eq \ref{giav}.

(ii) if $a_M > \alpha_{q=1} $ i.e. $f(a_M) =\delta > D(1)=f( \alpha_{q=1}) =\alpha_{q=1}  $,
then the saddle-point calculus in Eq. \ref{gsquarem} takes place at the boundary $\alpha_{min}$
and leads to
\begin{eqnarray}
 G^I &&  \simeq    \pi \rho(E)  N^{f(\alpha_{min})-\alpha_{min}} 
\label{gdommini}
\end{eqnarray}
so that here the statistics of $G^I$
is determined by the probability distribution of $\alpha_{min}$.
Let us again make the change of variables as in Eq. \ref{bmxi}
\begin{eqnarray}
 \alpha_{min}=a_M+b_M \xi
\label{bmxibis}
\end{eqnarray}
to find the appropriate scale $b_M$ that allows to obtain a rescaled random variable $\xi$ of order $O(1)$.
The linearisation
\begin{eqnarray}
f(\alpha_{min})=f(a_M+b_M \xi) \simeq f(a_M)+b_M \xi  f'(a_M) +... = \delta +b_M \xi  f'(a_M) +... 
\label{dvlin}
\end{eqnarray}
yields for the cumulative distribution of Eq. \ref{cumulalphamind}
\begin{eqnarray}
Prob( \alpha_{min}= a_M+b_M \xi \leq {\rm min}(\alpha_1,...,\alpha_M)  ) && 
\simeq  e^{-  c \rho(E) N^{f(a_M+b_M \xi)-\delta}      } \simeq e^{-  c \rho(E) e^{\xi b_M   f'(a_M) \ln N }}
\label{cumulalphamindxi}
\end{eqnarray}
The scale $b_M$ is thus given by
\begin{eqnarray}
 b_M =\frac{1}{  f'(a_M) \ln N }
\label{bmd}
\end{eqnarray}
Its decay as $1/(\ln N)$ shows that the linearization of Eq. \ref{dvlin} is appropriate.
The limiting distribution for the rescaled variable $\xi$ is then 
again the Gumbel distribution \cite{extreme}
\begin{eqnarray}
Prob( \alpha_{min}= a_M+b_M \xi \leq {\rm min}(\alpha_1,...,\alpha_M)  ) && 
\simeq   e^{-  c \rho(E) e^{\xi }}
\label{gumbeld}
\end{eqnarray}
Plugging Eq. \ref{bmxi} into Eq. \ref{gdommini} yields 
\begin{eqnarray}
 G^I &&  \simeq    \pi \rho(E)  N^{ f(a_M+b_M \xi)- (a_M+b_M \xi)} \simeq  \pi \rho(E)  N^{ f(a_M)+b_M \xi f'(a_M)- (a_M+b_M \xi)} 
\nonumber \\
&&  \simeq \pi \rho(E)  N^{ \delta-a_M} e^{ \xi \ln N b_M ( f'(a_M)-1)  }  \simeq \pi \rho(E)  N^{ \delta-a_M} e^{ - \xi  (\frac{1}{f'(a_M)}-1 )  } 
\label{gdomminid}
\end{eqnarray}
It is thus convenient to introduce 
the typical value corresponding to the value at $\xi=0$
\begin{eqnarray}
 G^I_{typ} && =\pi \rho(E)  N^{ \delta-a_M}
\label{gtypd}
\end{eqnarray}
and the index
\begin{eqnarray}
\mu \equiv \frac{ f'(a_M) }{1- f'(a_M) }
\label{mud}
\end{eqnarray}
where $a_M$ is the solution of $f(a_M)=\delta$ (Eq. \ref{eqforam}) :
as a consequence, the index $\mu$ depends on $\delta$ and on the multifractal spectrum $f(\alpha)$ but does not depend on $N$ (in contrast to 
the index $\mu_N$ of Eq. \ref{musmall} in the previous section).
In the limiting case $\delta \to 1^-$ where $a_M \to \alpha_0^-$
and $f'(a_M) \to 0$, one obtains $\mu \to 0$
as it should to match the results of the previous section. 
In the other limiting case $\delta \to D(1)=\alpha_{q=1}=f(\alpha_{q=1})$
 where $a_M \to \alpha_{q=1}$ and $f'(a_M) \to 1$, one obtains $\mu \to +\infty$.

 Eq. \ref{gdomminid} then reads
\begin{eqnarray}
 G^I &&  \simeq   G^I_{typ} e^{ - \frac{ \xi}{\mu} }
\label{gdomminidres}
\end{eqnarray}
so that the Gumbel probability distribution for $\xi$ (Eq. \ref{gumbeld})
\begin{eqnarray}
g(\xi) = - \frac{d}{d\xi}   e^{-  c \rho(E) e^{\xi }}
\simeq   c \rho(E) e^{ \xi -  c \rho(E) e^{\xi }}
\label{gumbeldxi}
\end{eqnarray}
transforms into the Fr\'echet distribution of index $\mu$ for $G^I$
\begin{eqnarray}
P( G^I ) && = \int_{-\infty}^{+\infty} d\xi g(\xi) \delta\left( G^I - G^I_{typ}
  e^{- \frac{\xi}{\mu} } \right)
\nonumber \\
&& = \frac{ c \rho(E) \mu  (G^I_{typ} )^{\mu} }{ (G^I)^{1+\mu} }
 e^{- c \rho(E) \left( \frac{G^I_{typ}}{G^I} \right)^{\mu} }
\label{frechetd}
\end{eqnarray}

As a consequence, the contribution of the typical region
to the moments of indices $q<\mu$ read
\begin{eqnarray}
\left[\int dG^I P( G^I ) (G^I)^q \right]^{\rm typical \  region}_{ q < \mu}  
 && = [c \rho(E) ]^{\frac{q}{\mu}} (G^I_{typ})^q  \int_0^{+\infty} dt e^{- t }  t^{- \frac{q}{\mu}}
 \nonumber \\
 && =  [c \rho(E) ]^{\frac{q}{\mu}} (G^I_{typ})^q \Gamma\left(1-\frac{q}{\mu} \right)
\label{gqsmall}
\end{eqnarray}
so that the scaling with the size $N$ directly reflects the scaling of 
the typical value $G^I_{typ} $ of Eq. \ref{gtypd}
\begin{eqnarray}
\left[\int dG^I P( G^I ) (G^I)^q \right]^{\rm typical region}_{ q < \mu}  
 && \propto (G^I_{typ})^q \propto  N^{(\delta -a_M) q}
\label{gqsmallscaling}
\end{eqnarray}

The moments $q>\mu$ that do not exist for the probability distribution in the scaling region ( Eq. \ref{frechetd})
are dominated by the rare events outside this scaling region, and should be thus evaluated 
from the statistics of $\alpha_{min}$ outside the scaling region of Eq. \ref{gumbeld} as we now describe. These rare events  will also give contributions for the moments $q<\mu$ that should be compared with Eq. \ref{gqsmallscaling} to decide
what is the biggest contribution between the two.

\subsection{ Multifractality analysis in the rare event region  }

As explained around Eq \ref{minlargedev}, the probability distribution of $\alpha_{min}$ 
in the region where $\alpha_{min}$ is anomalously small $\alpha_{min} \ll a_M$ reads
\begin{eqnarray}
P_{min}( \alpha_{min}  ) 
&& \opsimeq_{\alpha_{min} \ll a_M} M  \pi_N(\alpha_{min} ) \simeq c \rho(E) N^{1-\delta}  N^{f(\alpha_{min})-1} = c \rho(E) N^{f(\alpha_{min})-\delta} 
\label{minlargedevd}
\end{eqnarray}
while the corresponding
 the imaginary Green function is given by (Eq. \ref{gianomalous}) 
\begin{eqnarray}
 G^I && \opsimeq_{G^I \gg G^I_{typ}}  \frac{ \pi }{\eta}  N^{-\alpha_{min}} \simeq \frac{ \pi   }{ c }  N^{\delta-\alpha_{min}}
\label{gianomalousd}
\end{eqnarray}

The contribution of this rare-event region to the moment of order $q$ reads
\begin{eqnarray}
\left[ \overline{ ( G^I)^q } \right]^{\rm rare region} && \simeq   \int d\alpha_{min}  c \rho(E) N^{f(\alpha_{min})-\delta} 
\left[ \frac{ \pi   }{ c }  N^{\delta-\alpha_{min}}  \right]^q
\nonumber \\
&&  \simeq  \pi^q c^{1-q}  \rho(E)  N^{ (q-1)\delta}  \int d\alpha_{min}   N^{f(\alpha_{min}) -q \alpha_{min}} 
\nonumber \\
&&  \simeq  \pi^q c^{1-q}  \rho(E)  N^{ (q-1)\delta} Y_q(N)
\label{gmultifqd}
\end{eqnarray}

In terms of the generalized dimensions $D(q)$ or the anomalous dimension $\Delta(q)$ of Eq. \ref{Deltaqtauq}
\begin{eqnarray}
Y_q(N) \propto N^{(1-q) D(q) } = N^{1-q-\Delta(q)}
\label{yqdqdeltaq}
\end{eqnarray}
the dependence with respect to the size $N$ thus read
\begin{eqnarray}
\left[ \overline{ ( G^I)^q } \right]^{\rm rare \  region} && \propto   N^{(1-q) ( D(q)-\delta)  }
\nonumber \\
&&  \propto   N^{(1-q)(1-\delta) -\Delta(q)}
\label{gmultifrare}
\end{eqnarray}

\subsection{ Summary on the results in the non-standard scaling regime $0 \leq \delta <1$ }

\label{sec_summary}

Let us summarize the behaviors obtained in the present section for
 the non-standard scaling regime $0 \leq \delta <1$ :

(i) in the region $\delta<D(1)$, the imaginary Green function is expected to be self-averaging (Eq. \ref{gversavbis}).

(ii) in the region $\delta>D(1)$, the imaginary Green function 
displays the following multifractal properties 
in terms of the index $\mu$ of Eq. \ref{mud}

(ii-a) the moments $q>\mu$ are governed by the rare event scaling of Eq. \ref{gmultifrare}
\begin{eqnarray}
\left[ \overline{ ( G^I)^q } \right]_{q>\mu} && \propto   N^{(1-q) ( D(q)-\delta)  }
  =   N^{(1-q)(1-\delta) -\Delta(q)}
\label{gmultifdbig}
\end{eqnarray}

(ii-b) the moments $q<\mu$ are governed by the biggest contribution between
the contribution of the scaling region (Eq. \ref{gqsmallscaling})
and the contribution of rare event region (Eq. \ref{gmultifrare})
\begin{eqnarray}
\left[ \overline{ ( G^I)^q }  \right]_{ q < \mu}  
 && \propto {\rm max} [  N^{(\delta -a_M) q} ,   N^{(1-q) ( D(q)-\delta)  } ]
\label{gqsmallmax}
\end{eqnarray}

Let us now describe the explicit results for two simples cases.

\subsection{ Application to the Gaussian multifractal spectrum}

For the Gaussian multifractal spectrum described in section \ref{sec_lognormal}
with $1<\alpha_0<2$, the information dimension reads (Eq. \ref{dqgauss}) 
\begin{eqnarray}
D(1)= 2-\alpha_0 
\label{d1gauss}
\end{eqnarray}

The solution of $f(a_M)=\delta$ (Eq. \ref{eqforam}) yields
\begin{eqnarray}
a_M=\alpha_0- 2 \sqrt{(\alpha_0-1)(1-\delta)}
\label{amgauss}
\end{eqnarray}
The derivative 
\begin{eqnarray}
f'(a_M)= \sqrt{ \frac{1-\delta}{\alpha_0-1}}
\label{fpamgauss}
\end{eqnarray}
yields the index $\mu$ of Eq. \ref{mud} 
\begin{eqnarray}
\mu = \frac{ 1 }{\sqrt{ \frac{\alpha_0-1}{1-\delta}} -1 }
\label{mudgauss}
\end{eqnarray}

The summary of section \ref{sec_summary} becomes

(i) for $0<\delta<D(1)=2-\alpha_0$, the imaginary Green function is self-averaging.

(ii) for $D(1)<\delta<1$, the imaginary Green function 
displays multifractal properties 
in terms of the index $\mu$ of Eq. \ref{mud} :

(ii-a) the moments $q>\mu$ are governed by Eq. \ref{gmultifdbig} using $D(q)$ of Eq. \ref{dqgauss}
\begin{eqnarray}
\left[ \overline{ ( G^I)^q } \right]_{q>\mu} && \propto   N^{(1-q) ( 1-\delta-(\alpha_0-1)q )  }
\label{gmultifdbiggauss}
\end{eqnarray}

(ii-b) the moments $q<\mu$ are governed by Eq. \ref{gqsmallmax}
\begin{eqnarray}
\left[\int dG^I P( G^I ) (G^I)^q \right]_{ q < \mu}  
 && \propto {\rm max} [  N^{(\delta -\alpha_0+ 2 \sqrt{(\alpha_0-1)(1-\delta)} ) q} , 
 N^{(1-q) ( 1-\delta-(\alpha_0-1)q )  } ]
\label{gqsmallmaxgauss}
\end{eqnarray}

\subsection{ Application to the linear multifractal spectrum}

For the linear Gaussian multifractal spectrum described in section \ref{sec_levy}
with $0<\nu<1$, the information dimension vanishes (Eq. \ref{dqlevy}) 
\begin{eqnarray}
D(1)=0
\label{d1levy}
\end{eqnarray}

The solution of $f(a_M)=\delta$ (Eq. \ref{eqforam}) yields
\begin{eqnarray}
a_M= \frac{\delta}{\nu}
\label{amlevy}
\end{eqnarray}
The derivative 
\begin{eqnarray}
f'(a_M)= \nu
\label{fpamlevy}
\end{eqnarray}
yields the index $\mu$ of Eq. \ref{mud} 
\begin{eqnarray}
\mu = \frac{ \nu }{1-\nu } >\nu
\label{mudlevy}
\end{eqnarray}

The summary of section \ref{sec_summary} becomes

(i) the self-averaging region $\delta<D(1)=0$ does not exist.

(ii) for $0<\delta<1$, the multifractal properties are as follows :

(ii-a) for $q>\mu = \frac{ \nu }{1-\nu }$, Eq. \ref{gmultifdbig} reads using $D(q>\mu>\nu)=0$
(Eq \ref{dqlevy})
\begin{eqnarray}
\left[ \overline{ ( G^I)^q } \right]_{q>\mu} && \propto   N^{(1-q) ( D(q)-\delta)  } = N^{(q-1) \delta}
\label{gmultifdbiglevy}
\end{eqnarray}

(ii-b) for $q<\mu$, Eq. \ref{gqsmallmax} reads
\begin{eqnarray}
\left[\int dG^I P( G^I ) (G^I)^q \right]_{ q < \mu}  
 && \propto {\rm max} [  N^{ -\frac{\delta}{\mu}  q} ,   N^{(1-q) ( D(q)-\delta)  } ]
\label{gqsmallmaxlevy}
\end{eqnarray}
Since $D(q)$ changes at $q=\nu$ (Eq \ref{dqlevy}), one needs to distinguish further the two regions : 

- for $\nu<q<\mu$ where $D(q)=0$, Eq. \ref{gqsmallmaxlevy} yields
\begin{eqnarray}
\left[\int dG^I P( G^I ) (G^I)^q \right]_{\nu< q < \mu}  
 && \propto {\rm max} [  N^{ -\frac{\delta}{\mu} q} ,   N^{(q-1)\delta } ] =  N^{(q-1)\delta }
\label{gqsmallmaxlevybiggernu}
\end{eqnarray}

- for $q<\nu$, using Eq. \ref{dqlevy}, Eq. \ref{gqsmallmaxlevy} becomes
\begin{eqnarray}
\left[\int dG^I P( G^I ) (G^I)^q \right]_{ q < \nu}  
 && \propto {\rm max} [  N^{ -\frac{\delta}{\mu}  q} ,   N^{1-\frac{q}{\nu} +(q-1)\delta  } ]
= N^{1-\frac{q}{\nu} +(q-1)\delta  }
\label{gqsmallmaxlevysmallernu}
\end{eqnarray}

In conclusion, all the moments can be actually summarized by the single
 rare event formula of Eq. \ref{gmultifrare}
\begin{eqnarray}
\left[\int dG^I P( G^I ) (G^I)^q \right] 
 && \propto N^{(1-q) (D(q)-\delta)  }
\label{gqsummarylevy}
\end{eqnarray}
in terms of generalized dimension $D(q)$ of Eq. \ref{dqlevy}
 that changes of behavior at $q=\nu$.

\section{ Conclusion}

\label{sec_conclusion}

In this paper, we have considered Anderson Localization models defined on $N$ sites :
we have assumed the multifractality of eigenvectors, and we have studied the consequences
for the statistical properties of the Green function $G_{xy}(E-i \eta)$.

After the analysis of the heavy-tails appearing in the probability of the real Green function $G^R_{xy}(E)$ for $\eta=0$,
we have focused on the statistical properties of the imaginary Green function $G^I_{xx}(E-i \eta)$
depending on the scaling relation between the broadening $\eta=\frac{c}{N^{\delta}}$ and the finite size $N$ :

(a) For the standard scaling $\delta=1$, we have analyzed the two regimes $c \ll 1$ and $c \gg 1$,
and we have found that the moments of the imaginary Green function are governed by the anomalous exponents $\Delta(q)$ in both cases, even if the probability distributions are very different.
Our conclusion is thus that the standard equivalence of Eq. \ref{ldos1state} is valid without the assumption of 
strong correlations between the weights.

(b) For the non-standard scaling $0<\delta<1$, we have derived the Fr\'echet probability distribution for the imaginary Green function
in the typical region and analyzed the multifractal properties of the moments. We have described the application to
the case of Gaussian multifractality and to the case of linear multifractality.

\end{document}